# Observation of many-body scarring in a Bose-Hubbard quantum simulator

Guo-Xian Su,[1,2,3] Hui Sun,[1,2,3] Ana Hudomal,[4,5] Jean-Yves Desaules,[4] Zhao-Yu Zhou,[1,2,3] Bing Yang,[6]
Jad C. Halimeh,[7] Zhen-Sheng Yuan,[1,2,3,*] Zlatko Papić,[4,†] and Jian-Wei Pan[1,2,3,‡]

[1]*Hefei National Laboratory for Physical Sciences at Microscale and Department of Modern Physics,
University of Science and Technology of China, Hefei, Anhui 230026, China*
[2]*Physikalisches Institut, Ruprecht-Karls-Universität Heidelberg, Im Neuenheimer Feld 226, 69120 Heidelberg, Germany*
[3]*CAS Center for Excellence and Synergetic Innovation Center in Quantum Information and Quantum Physics,
University of Science and Technology of China, Hefei, Anhui 230026, China*
[4]*School of Physics and Astronomy, University of Leeds, Leeds LS2 9JT, United Kingdom*
[5]*Institute of Physics Belgrade, University of Belgrade, 11080 Belgrade, Serbia*
[6]*Department of Physics, Southern University of Science and Technology, Shenzhen 518055, China*
[7]*INO-CNR BEC Center and Department of Physics, University of Trento, Via Sommarive 14, I-38123 Trento, Italy*

The ongoing quest for understanding nonequilibrium dynamics of complex quantum systems underpins the foundation of statistical physics as well as the development of quantum technology. Quantum many-body scarring has recently opened a window into novel mechanisms for delaying the onset of thermalization by preparing the system in special initial states, such as the $\mathbb{Z}_2$ state in a Rydberg atom system. Here we realize many-body scarring in a Bose-Hubbard quantum simulator from previously unknown initial conditions such as the unit-filling state. We develop a quantum-interference protocol for measuring the entanglement entropy and demonstrate that scarring traps the many-body system in a low-entropy subspace. Our work makes the resource of scarring accessible to a broad class of ultracold-atom experiments, and it allows one to explore the relation of scarring to constrained dynamics in lattice gauge theories, Hilbert space fragmentation, and disorder-free localization.

## I. INTRODUCTION

Coherent manipulation of quantum many-body systems far from equilibrium is key to unlocking outstanding problems in quantum sciences including strongly coupled quantum field theories, exotic phases of matter, and development of enhanced metrology and computation schemes. These efforts, however, are frequently plagued by the presence of interactions in such systems, which lead to fast thermalization and information scrambling: the behavior known as quantum ergodicity [1–3]. A twist came with recent advances in synthetic quantum matter, which enabled detailed experimental study of thermalization dynamics in isolated quantum many-body systems, leading to the observation of ergodicity-violating phenomena in integrable [4] and many-body localized systems [5,6].

More recently, quantum many-body scarring has emerged as another remarkable ergodicity-breaking phenomenon, where preparing the system in special initial states effectively traps it in a "cold" subspace that does not mix with the thermalizing bulk of the spectrum [7,8]. Such behavior hinders the scrambling of information encoded in the initial state and suppresses the spreading of quantum entanglement, allowing a many-body system to display persistent quantum revivals. Many-body scarring was first observed in the Rydberg atom experimental platform [9,10], and subsequent observations of weak ergodicity breaking phenomena have attracted much attention [11–13]. On the other hand, theoretical works have unearthed universal scarring mechanisms [14–17], pointing to the ubiquity of scarring phenomena in periodically driven systems [18–20] and in the presence of disorder [21,22]. Given that many-body scarring in Rydberg atom systems has previously been reported in a single initial state, the $\mathbb{Z}_2$-ordered state, many questions remain about the overall fragility of this phenomenon and its sensitivity to the initial condition. It is thus vital to extend the realm of scarring to a greater variety of experimental platforms and more accessible initial conditions, which would empower a fundamental understanding of nonergodic dynamics in various research areas ranging from lattice gauge theories to constrained glassy systems.

In this paper, we observe many-body scarring in a large-scale Bose-Hubbard quantum simulator, where we employ a tilted optical lattice to emulate the PXP model, a canonical model of many-body scarring [23–26]. We demonstrate that many-body scarring can result from a larger set of initial states, including the unit-filling state at finite detuning, hitherto believed to undergo fast thermalization [9]. Furthermore, we demonstrate that periodic driving can be used to enhance scarring behavior. Taking advantage of spin-dependent optical superlattices, we measure the system's entanglement entropy by interfering identical copies in the double wells. We show the average entropy of single-site subsystems to be a good approximation of half-chain bipartite entropy, revealing a key property of scarring: the "trapping" of the quantum system in a low-entropy subspace, which prevents its relaxation into the exponentially large Hilbert space.

*yuanzs@ustc.edu.cn
†Z.Papic@leeds.ac.uk
‡pan@ustc.edu.cn

The remainder of this paper is organized as follows. In Sec. II we introduce our experimental setup and show how it can realize the PXP model. In Sec. III we benchmark our quantum simulation by observing many-body scarring from the previously known $\mathbb{Z}_2$ initial state. We also demonstrate the enhancement of scarring under periodic driving. In Sec. IV we present our measurements of entanglement entropy, providing deeper insight into the slow thermalization dynamics associated with scarred initial states. Finally, in Sec. V we extend the scarring phenomenon to a regime at moderate detuning for the unit-filling initial state. Our conclusions are presented in Sec. VI, while Appendixes A–D contain the derivation of the PXP mapping, further details on state preparation and measurement techniques, and a numerical study of other scarred initial conditions.

## II. MAPPING THE PXP MODEL ONTO THE BOSE-HUBBARD MODEL

The PXP model [27,28] describes a kinetically constrained chain of spin-1/2 degrees of freedom. Each spin can exist in two possible states, $|\circ\rangle$ and $|\bullet\rangle$, corresponding to the ground state and excited state, respectively. An array of $N$ such spins is governed by the Hamiltonian

$$\hat{H}_{\text{PXP}} = \Omega \sum_{j=1}^{N} \hat{P}_{j-1} \hat{X}_j \hat{P}_{j+1}, \quad (1)$$

where $\hat{X} = |\circ\rangle\langle\bullet| + |\bullet\rangle\langle\circ|$ is the Pauli $x$ matrix, describing local spin precession with frequency $\Omega$. The projectors onto the ground state, $\hat{P} = |\circ\rangle\langle\circ|$, constrain the dynamics by allowing a spin to flip only if both of its neighbors are in the ground state.

A remarkable property of the PXP model is that it is quantum chaotic and yet it exhibits persistent quantum revivals from a highly out-of-equilibrium $|\mathbb{Z}_2\rangle \equiv |\bullet\circ\bullet\circ\cdots\rangle$ initial state [23,29–31]. The presence of revivals from a special initial state in an overall chaotic system was understood to be a many-body analog of the phenomena associated with a single particle inside a stadium billiard, where nonergodicity arises as a "scar" imprinted by a particle's classical periodic orbit [16,32,33]. In many-body scarred systems, eigenstates were shown to form tower structures [23]. These towers are revealed by the anomalously high overlap of eigenstates with special initial states, and their equal energy spacing is responsible for quantum revivals. While previous experiments on Rydberg atoms [9,10] have primarily focused on the $|\mathbb{Z}_2\rangle$ initial state, we will demonstrate that the PXP model can effectively emerge in the Bose-Hubbard model, allowing us to identify scarred revivals from a larger set of initial conditions, including the polarized state $|0\rangle \equiv |\circ\circ\circ\cdots\rangle$.

Our experiment begins with a $^{87}$Rb Bose-Einstein condensate, which is compressed in the $z$ direction and loaded into a single layer of a pancake-shaped trap. We then perform the superfluid-to-Mott-insulator phase transition with optical lattices in the $x$-$y$ plane. In both $x$ and $y$ directions, we have a superlattice that is formed by superimposing the "short" lattice, with $a_s = 383.5$ nm spacing, and the "long" lattice, with $a_l = 767$ nm spacing [34,35], each of which can be individually controlled. We realize independent one-dimensional (1D) Bose-Hubbard systems in the $y$ direction by ramping up the short-lattice depth in the $x$ direction over $40E_r$, with $E_r = h^2/8ma_s^2$ being the short-lattice recoil energy, where $h$ is the Planck constant and $m$ is the $^{87}$Rb atomic mass. The short lattice in the $y$ direction makes an approximately 4° angle with gravity, which results in a static linear tilt per site of $\Delta_g = 816$ Hz; see Fig. 1(a). An external magnetic field gradient $\Delta_B$ may be further added to create a tunable linear tilting potential $\Delta = \Delta_g + \Delta_B$. The effective Hamiltonian describing our simulator is

$$\hat{H} = -J \sum_{i=1}^{L-1} (\hat{b}_i^\dagger \hat{b}_{i+1} + \hat{b}_{i+1}^\dagger \hat{b}_i) + \hat{U} + \hat{\Delta}, \quad (2)$$

where $J$ is the hopping amplitude, $\hat{b}$ and $\hat{b}^\dagger$ are the standard Bose annihilation and creation operators, the interaction energy is $\hat{U} = (U/2) \sum_{i=1}^{L} \hat{n}_i(\hat{n}_i - 1)$, and the tilt potential is $\hat{\Delta} = \Delta \sum_{i=1}^{L} i\hat{n}_i$. $L$ denotes the number of sites in the chain with open boundary conditions, and we restrict ourselves to a total filling equal to 1, i.e., with the same number of bosons as lattice sites.

In order to realize the PXP model in the Bose-Hubbard quantum simulator, we tune the parameters to the resonant regime $U \approx \Delta \gg J$ [36,37], which has been studied extensively in the context of quantum Ising chains [38–40]. In this regime, three-boson occupancy of any site is strongly suppressed, and doublons can only be created by moving a particle to the left, e.g., $\cdots 11 \cdots \to \cdots 20 \cdots$, or destroyed by moving a particle to the right. The states of the PXP model are understood to live on the bonds of the Bose-Hubbard model. An excitation in the PXP model $\bullet_{j,j+1}$, living on the bond $(j, j+1)$, corresponds to the creation of a doublon $2_j 0_{j+1}$ on site $j$ in the Bose-Hubbard chain. We identify the unit-filling state $|111\cdots\rangle$ with the PXP polarized state, $|0\rangle$. Any other configuration of the PXP model can be mapped to a Fock state in the Bose-Hubbard model by starting from the unit filling, identifying the bonds that carry PXP excitations and replacing the corresponding sites in the Mott state with $11 \to 20$. Applying this rule across the chain allows us to map any basis state of the PXP model to a corresponding Fock state in the Bose-Hubbard model; for example, the $|\mathbb{Z}_2\rangle$ state maps to the Fock state $|\cdots 2020 \cdots\rangle$. Figure 1(b) illustrates the profound change in the connectivity of the Fock space near the resonance $U \approx \Delta \gg J$, with an emergent dynamical subspace isomorphic to the PXP model in the sector containing the $|\mathbb{Z}_2\rangle$ state. For a detailed derivation of the mapping, see Appendix A.

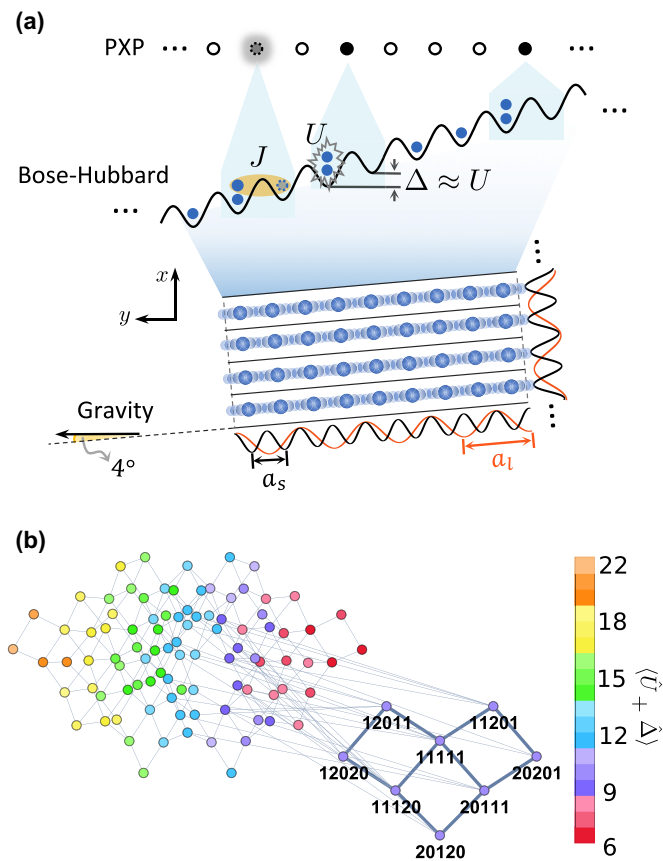

FIG. 1. Realizing the PXP model in a Bose-Hubbard quantum simulator. (a) A schematic of the optical lattice. Deep lattice potential in the $x$ direction forms isolated chains in the $y$ direction, where the linear tilting potential is applied. Spin-dependent superlattices consisting of two standing waves in each direction can be individually controlled for state preparation and measurement. At the resonance $U \approx \Delta \gg J$, the dominant hopping process is $11 \leftrightarrow 20$. The PXP excitations, ●, live on the bonds between the lattice sites. The doublon configuration 20 in the Bose-Hubbard model maps to an excitation in the PXP model, while all other configurations are mapped to an empty site, ○. For example, the given state $|\cdots\circ\bullet\circ\bullet\circ\circ\circ\bullet\cdots\rangle$ maps to the Fock state $|\cdots 120201120\cdots\rangle$. (b) Emergence of the PXP subspace in the Bose-Hubbard model at the resonance $U \approx \Delta \gg J$. Dots represent Fock states of the tilted Bose-Hubbard model with five bosons on five sites (restricting ourselves to at most three bosons on any site). Lines denote the allowed hopping processes. The color scale shows the sum of interaction and tilt energies $\langle \hat{U} + \hat{\Delta} \rangle$ for each Fock state, and this value is conserved by resonant processes. The PXP dynamical subspace and its Fock states are explicitly labeled.

### III. OBSERVATION OF $\mathbb{Z}_2$ QUANTUM MANY-BODY SCARS

To prepare the initial states, we first employ an entropy redistribution cooling method [34] with the superlattice in the $y$ direction to prepare an $\bar{n} = 2$ Mott insulator in the left (odd) sites, while removing all atoms on the right (even) sites via site-dependent addressing [35]. This gives us the initial state $|\psi_0\rangle = |\mathbb{Z}_2\rangle = |2020\cdots\rangle$ (see Appendix B). In the region of interest, we have prepared 50 copies of the initial state $|\psi_0\rangle$ isolated by the short lattice along the $x$ direction. Each copy extends over 50 short lattice sites along the $y$ direction.

We quench the system out of equilibrium by abruptly dropping the $y$-lattice depth to $11.6E_r$, which corresponds to switching $J$ from 0 to 51(1) Hz. This is done while simultaneously adjusting the lattice depth in the $x$ and $z$ directions accordingly, such that the interaction strength matches the linear tilt with $U = \Delta = \Delta_g \approx 16J$. After evolution time $t$, we freeze the dynamics by ramping up the $y$-lattice depth rapidly and read out the atomic density on the left ($\langle \hat{n}_\text{Left}\rangle$) and right ($\langle \hat{n}_\text{Right}\rangle$) sites of the double wells formed by the $y$ superlattice successively [35,41]. This provides access to the density imbalance, $\langle \hat{M}_z \rangle = (\langle \hat{n}_\text{Left}\rangle - \langle \hat{n}_\text{Right}\rangle)/(\langle \hat{n}_\text{Left}\rangle + \langle \hat{n}_\text{Right}\rangle)$, an observable corresponding to the staggered magnetization in the PXP model; see Fig. 2(a). Another observable is the density of excitations in the PXP model, which is measured by projecting out the even atomic number occupancy on each site, then reading out the average odd particle density $\langle \hat{P}_{\hat{n}\in\text{odd}}\rangle_{(1)}$ [41]. Due to highly suppressed multiboson occupancy, we have $\langle \hat{P}_{|\bullet\rangle}\rangle = \langle \hat{n}_\text{doublon}\rangle_{(1)} \approx (1 - \langle \hat{P}_{\hat{n}\in\text{odd}}\rangle_{(1)})/2$.

Away from the resonance, the dynamics is ergodic, and the staggered magnetization present in the initial $|\mathbb{Z}_2\rangle$ state quickly decays with time; see Fig. 2(b). In contrast, by tuning to the vicinity of the resonance, $\Delta = U$, we observe distinct signatures of scarring: The system approximately undergoes persistent oscillations between the $|\mathbb{Z}_2\rangle \equiv |\bullet\circ\bullet\circ\cdots\rangle$ configuration and its partner shifted by one site, $|\bar{\mathbb{Z}}_2\rangle \equiv |\circ\bullet\circ\bullet\cdots\rangle$, as can be seen in the staggered magnetization profile and the density of excitations in Fig. 2(b). The density of excitations does not distinguish between $|\mathbb{Z}_2\rangle$ and $|\bar{\mathbb{Z}}_2\rangle$ states; hence there is a trivial factor of 2 difference between the oscillation frequencies of $\langle \hat{P}_{|\bullet\rangle}\rangle$ and $\langle \hat{M}_z\rangle$.

The scarred oscillations in Fig. 2(b) are visibly damped with a decay time $\tau = 49.6 \pm 0.8$ ms. Nevertheless, as shown in Ref. [10], by periodically driving the system it is possible to "refocus" the spreading of the many-body wave function in the Hilbert space and thereby enhance the scarring effect, as we demonstrate numerically in Fig. 2(c) and experimentally in Fig. 2(d). Our driving protocol is based on modulating the laser intensity of the $z$ lattice, which translates into periodic modulation of the interaction energy, $U(t) = \Delta + U_0 + U_m \cos(\omega t)$, while $\Delta$ is kept fixed. This results in a modulation of the density of doublons in the chain, acting as the analog of the chemical potential in the PXP model.

Numerical simulations of the PXP model with the driven chemical potential, shown in Fig. 2(c), demonstrate the dynamical stabilization of the Hilbert space trajectory. We visualize the trajectory by plotting the average sublattice occupations, $\langle \hat{n}_\text{Left}\rangle$ and $\langle \hat{n}_\text{Right}\rangle$, normalized to the interval [0,1]. The $|\mathbb{Z}_2\rangle$ and $|\bar{\mathbb{Z}}_2\rangle$ states are located at the coordinates (1,0) and (0,1), which are the lower right and upper left corners of this diagram, respectively. The polarized state $|0\rangle$ is at the origin (0,0).

In the undriven case [left panel of Fig. 2(c)], the trajectory at first oscillates between $|\mathbb{Z}_2\rangle$ and $|\bar{\mathbb{Z}}_2\rangle$ states, while passing through a region with a lower number of excitations. However, as the time passes, the trajectory drifts, exploring progressively larger parts of the Hilbert space. By contrast, when the driving is turned on [right panel of Fig. 2(c)], the

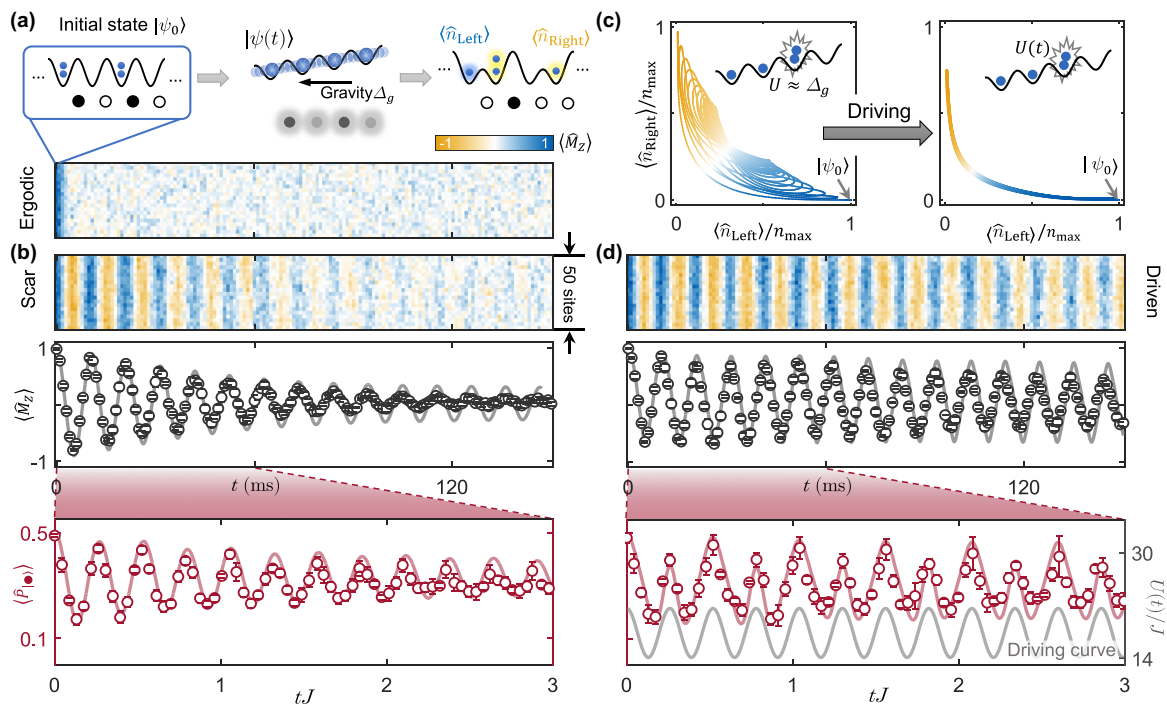

FIG. 2. Observation of $\mathbb{Z}_2$ quantum many-body scars in a Bose-Hubbard quantum simulator. (a) Starting from the state $|\psi_0\rangle = |\cdots 2020 \cdots\rangle$—the analog of the $|\mathbb{Z}_2\rangle$ state in the PXP model—we utilize gravity to provide linear tilt $\Delta = \Delta_g$. We characterize quench dynamics by measuring density imbalance and the number of doublons, corresponding to staggered magnetization $\langle \hat{M}_z \rangle$ and density of excitations $\langle \hat{P}_{|\bullet\rangle} \rangle$ in the PXP model. In the detuned regime $\Delta - U \approx -2J$, the dynamics is ergodic, and the system has no memory of the initial state at late times. (b) Tuning to $U \approx \Delta$, we observe persistent oscillations in both $\langle \hat{M}_z \rangle$ and $\langle \hat{P}_{|\bullet\rangle} \rangle$. This memory of the initial state is a signature of weak ergodicity breaking due to quantum many-body scars. (c) and (d) Periodic modulation of the interaction $U(t) = \Delta + U_0 + U_m \cos(\omega t)$ with $U_0 = 1.85J$, $U_m = 3.71J$, and $\omega = 3.85J$ leads to an enhancement of scarring. (c) shows the numerically computed trajectory in the sublattice occupation plane for the PXP model with $N = 24$ sites, with and without driving. The sublattice occupancies $\langle \hat{n}_{\text{Left}} \rangle$ and $\langle \hat{n}_{\text{Right}} \rangle$ are normalized to the interval [0,1]. The driving is seen to strongly suppress the spreading of the trajectory. In (d), experimental measurements on the driven Bose-Hubbard model show robust scarred oscillations at all accessible times. In both the static and driven cases, experimental data for $\langle \hat{M}_z \rangle$ and $\langle \hat{P}_{|\bullet\rangle} \rangle$ are in excellent agreement with TEBD numerical simulations shown by gray and red solid curves. The gray curve in the lowest panel shows the modulation $U(t)$.

trajectory approximately repeats the first revival period of the undriven case, even at late times. Thus the driving stabilized the scarred revivals without significantly altering their period.

Experimental measurements on the driven Bose-Hubbard model in Fig. 2(d) find a strong enhancement of the amplitude of the oscillations in staggered magnetization with the decay time $\tau$ increasing to $208 \pm 10$ ms, while the period remains nearly the same as in the static case. Optimal driving parameters were determined numerically using a combination of simulated annealing and brute-force search; see Supplemental Material [42].

We note that the experimental measurement of $\langle \hat{M}_z \rangle$ damps slightly faster than the theory prediction, shown by a curve in Fig. 2(b), at late times ($t > 60$ ms). We attribute this to an inherent residual inhomogeneity across the lattice, which results in dephasing between different parts of the system, as well as possible decoherence induced by scattering of the lattice lasers. To avoid the effect of these undesired dephasing or decoherence effects, in the following we limit our investigation to times up to 60 ms.

## IV. UNRAVELING THE DETAILS OF SCARRED DYNAMICS VIA QUANTUM INTERFERENCE

Entanglement entropy is key for characterizing scarring behavior. Entropy provides a window into the evolution of the system's wave function and the spreading of quantum entanglement. For a system trapped in a scarred subspace, thermalization is inhibited, and the system exhibits suppressed entropy growth and periodic fidelity revivals. Measuring these observables usually requires brute-force state tomography, but for our 50-site Bose-Hubbard system with a Hilbert space dimension exceeding $10^{28}$, this approach is generally impossible.

However, the superlattice in the $x$ direction allows us to probe entanglement entropy by interfering identical copies in the double wells, analogous to the 50 : 50 beam splitter (BS) interference employed in photonics experiments [43]; see Fig. 3(a). This is done by freezing the dynamics along the chains in the $y$ direction after evolution time $t$; then we interfere copies of $|\psi(t)\rangle$ in the double wells formed by the $x$ superlattice (see Appendix C). After the interference, a parity projection helps read out the average odd particle density $\langle \hat{P}^{\rm BS}_{\hat{n}\in \rm odd}\rangle_{(1)}$, which gives us access to the second-order Rényi entropy [44]. We measure the entropy of single-site subsystems $S_{(1)} = -\ln({\rm Tr}_{(1)}[\hat{\rho}(t)^2]) = -\ln(1 - 2\langle \hat{P}^{\rm BS}_{\hat{n}\in \rm odd}\rangle_{(1)})$, where $\hat{\rho}(t) = |\psi(t)\rangle \langle \psi(t)|$ is the density matrix. Entanglement entropy $S_{(1)}$, shown in Fig. 3(b), grows much more slowly than expected in a thermalizing system. The growth is accompanied by oscillations with the same frequency as $\langle \hat{P}_{|\bullet\rangle}\rangle$ in Fig. 2(b), implying that the system returns to the neighborhood of product states $|\mathbb{Z}_2\rangle$ and $|\bar{\mathbb{Z}}_2\rangle$. Furthermore, the entropy growth becomes almost fully suppressed by periodic driving, indicating that the scarred subspace disconnects from the thermalizing bulk of the spectrum. Numerical time-evolving block decimation (TEBD) simulations confirm that this lack of thermalization at the single-site level provides a good approximation for the behavior of larger subsystems, as demonstrated by the half-chain bipartite entropy $S_{L/2}$ plotted in Fig. 3(b). This shows that scarring traps the system in a vanishingly small corner of an exponentially large Hilbert space.

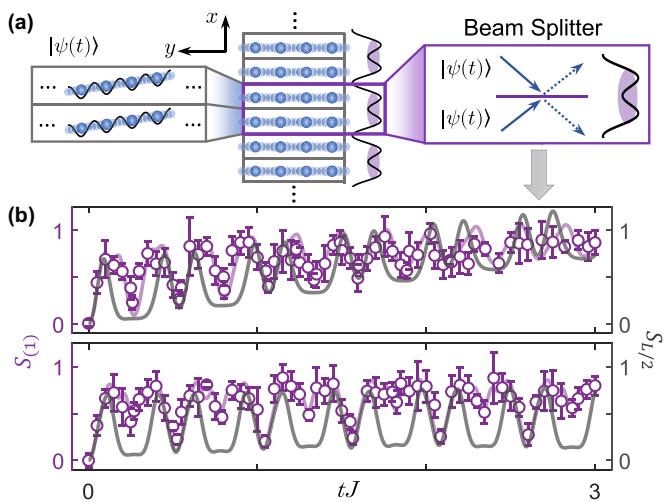

FIG. 3. Probing many-body scarred dynamics via quantum interference. (a) After evolution time $t$, we freeze the dynamics in the $y$ direction, and then by interfering two identical copies in the double wells along the $x$ direction, we obtain the second-order Rényi entropy. (b) The entropy for a single site, $S_{(1)}$, is seen to have robust oscillations with the same frequency as in Fig. 2(b), indicating a lack of thermalization. The slow growth of entropy in the absence of driving (upper panel) is strongly suppressed when we drive the system using the same parameters as in Fig. 2(d) (lower panel). In both cases, the single-site entropy is a good approximation to the half-chain entropy, $S_{L/2}$, evaluated numerically using TEBD (gray curve).

## V. EMERGENCE OF DETUNED SCARRING IN THE POLARIZED STATE

Up to this point, we have provided extensive benchmarks of our quantum simulator against the previously known case of $\mathbb{Z}_2$ quantum many-body scars [9]. In this section we demonstrate that our quantum simulator also hosts distinct scarring regimes for initial states other than $|\mathbb{Z}_2\rangle$, which are enabled by detuning and further stabilized by periodic drive. We highlight this finding by observation of scarring behavior in the polarized state $|0\rangle$, previously not associated with scars.

We first prepare the unit-filling state $|1111\cdots\rangle$ by transferring $|2, 0\rangle$ to $|1, 1\rangle$ states in the superlattice [34], which maps to the polarized state in the PXP model (see also Appendix B). In the absence of detuning or periodic drive, we observe fast relaxation: Both the density of excitations and single-site entropy rapidly relax, with no visible oscillations beyond the timescale $\sim 1/J$; see Fig. 4(a). Interestingly, when we bias the system by a static detuning, $U_0 = -2.38J$, we observe the emergence of oscillations in all three observables, accompanied by a slight decay; see Fig. 4(b). Finally, if we also periodically modulate the interaction with amplitude $U_m = 1.54J$ and frequency $\omega = 4.9J \times 2\pi$, we find a dramatic enhancement of scarring [Fig. 4(c)]. In particular, entropy now shows pronounced oscillations, signaling robust scar-induced coherence at all experimentally accessible times.

The intuitive picture behind our observations is summarized as follows. In the absence of detuning or periodic drive, the system initialized in the polarized state undergoes chaotic dynamics and rapidly explores the entire Hilbert space. By biasing the system via static detuning, thermalization can be suppressed over moderate timescales. Finally, by periodically driving the system it is possible to "refocus" the spreading of the many-body wave function in the Hilbert space and thereby enhance the scarring effect, similar to the findings of Ref. [10] for the $|\mathbb{Z}_2\rangle$ state. In the remainder of this section, we present our theoretical analysis of the experiment that supports this interpretation of the dynamics.

Figure 5 shows the results of exact diagonalizations of the PXP model in Eq. (1) in the presence of static detuning, $\hat{H}(\mu) = \hat{H}_{\rm PXP} + \mu_0 \sum_i \hat{n}_i$, where $\hat{n}_i$ takes a value equal to 1 if site $i$ contains an excitation and 0 otherwise. The static chemical potential $\mu_0$ is proportional to the Bose-Hubbard detuning parameter $U_0$ in Fig. 4. Figure 5(a) plots the overlap of all energy eigenstates $|E\rangle$ of the pure PXP model ($\mu_0 = 0$) with the polarized state $|\psi_0\rangle = |0\rangle$. As expected, we do not see any hallmarks of scars, such as ergodicity-violating eigenstates with anomalously enhanced projection on $|0\rangle$. Moreover, the lowest-entropy eigenstates, denoted by squares in Fig. 5(b), are the known $\mathbb{Z}_2$ scarred eigenstates [31] which are hidden in the bulk of spectrum when the overlap is taken with the $|0\rangle$ state.

On the other hand, when we add the static chemical potential $\mu_0 = 1.68\Omega$, corresponding to the detuning value in Fig. 4, a band of scarred eigenstates with anomalously large overlap with $|0\rangle$ emerges; see Fig. 5(c). The band of scarred eigenstates, highlighted by star symbols in Fig. 5(c), spans the entire energy spectrum, but their support on $|0\rangle$ is biased towards the ground state due to the breaking of particle-hole symmetry by detuning. The detuned scarred states also have anomalously low entanglement entropy, as seen in Fig. 5(d).

A few comments are in order. We note that exact diagonalization confirms that the PXP model remains chaotic for the value of detuning used in Fig. 5(c), and this detuning is not large enough to trivially fragment the entire spectrum into disconnected sectors with the given numbers of excitations [42]. Moreover, we confirmed that the scarred eigenstates in Fig. 5(c) are distinct from the ones associated with the $|\mathbb{Z}_2\rangle$ state in Fig. 5(a). Thus it remains to be understood if these eigenstates can be described within the su(2)

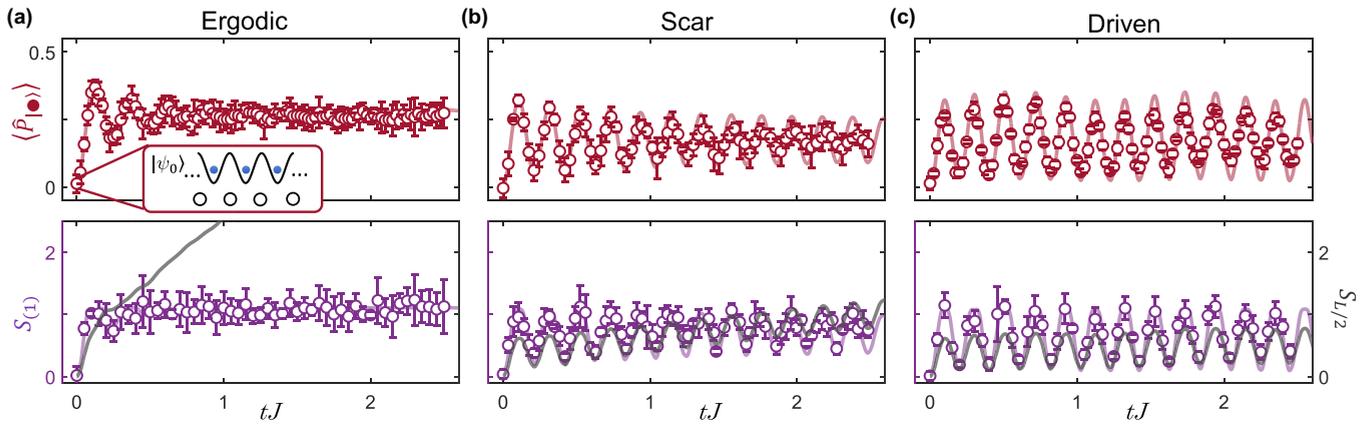

FIG. 4. Emergence of many-body scarring in the polarized state. (a) Fast thermalization from the unit-filling state in the Bose-Hubbard chain at $U = \Delta$ resonance. (b) Emergence of scarred dynamics in the presence of static detuning. (c) Dynamical stabilization of scarred dynamics in the presence of both detuning and periodic driving. The top and bottom rows show the experimental measurements of the density of excitations and second Rényi entropy, respectively. The ergodic case shows fast growth of the half-chain entropy compared with the single-site entropy, while for the scarred dynamics, the single-site entropy approximates well the half-chain entropy, with or without periodic driving. The static detuning is $U_0 = -2.38J$, and the modulation parameters are $U_m = 1.54J$ and $\omega = 4.9J \times 2\pi$. The curves are the results of TEBD simulations.

spectrum-generating algebra framework developed for the $|\mathbb{Z}_2\rangle$ state in Ref. [45].

Nevertheless, similar to the $|\mathbb{Z}_2\rangle$ case, the scarring from the $|0\rangle$ state can be further enhanced by periodic modulation of the PXP chemical potential, $\mu(t) = \mu_0 + \mu_m \cos(\omega t)$. By evaluating the corresponding Floquet operator, we find that a single Floquet mode develops a very large overlap with the $|0\rangle$ state [42]. The existence of a single Floquet mode, whose mixing with other modes is strongly suppressed, gives rise to robust oscillations in the dynamics well beyond the experimentally accessible timescales.

To probe the ergodicity of the dynamics from the polarized state, we compare the difference between the predictions of the diagonal and canonical ensembles for an observable such as the average number of excitations; see Fig. 6(a). These two ensembles are expected to give the same result if the strong eigenstate thermalization hypothesis (ETH) holds [46] and all eigenstates at a similar energy density yield the same expectation value for local observables. Figure 6(a) shows that the discrepancy between the two ensembles is the strongest around $\mu_0 \approx 1.68\Omega$, where we observe strong scarring. For $\mu_0/\Omega$ close to 0, the polarized state thermalizes quickly towards the thermal value expected for a state whose expectation value of the energy is near the middle of the many-body spectrum. For very large $\mu_0/\Omega$, we enter a trivial regime where the polarized state is close to the ground state and only a few eigenstates at low energies are relevant for the dynamics. Hence, in this regime, quenching from the polarized state is similar to quenching from a thermal state at a very low temperature, and the agreement between the two ensembles is again very good. However, in this regime, only a very small part of the many-body Hilbert space is explored by the dynamics. This is not the case in the scarred regime that we investigate experimentally, and this can be demonstrated by studying the relevant classical limit, as shown next.

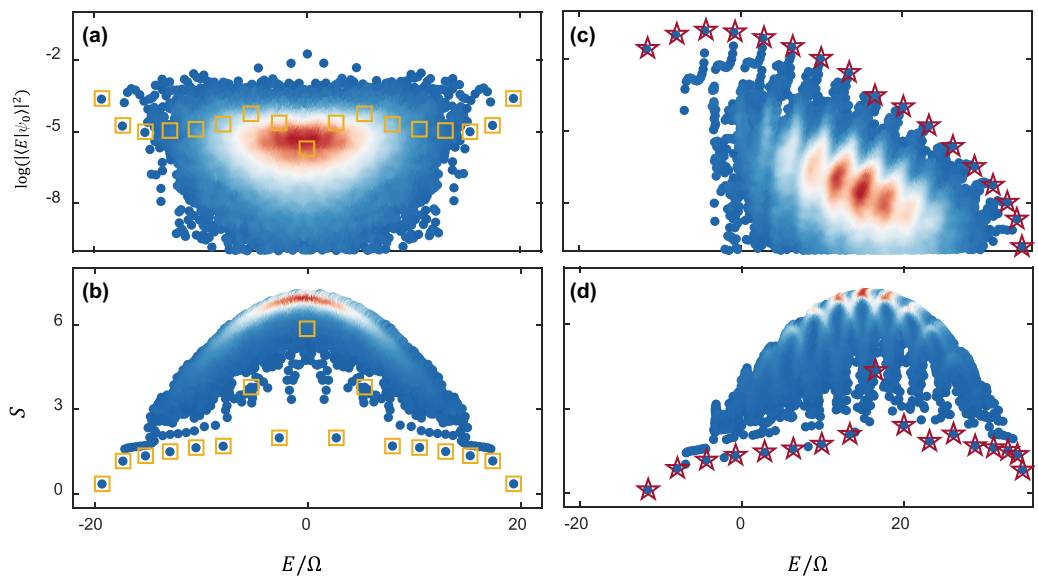

FIG. 5. Eigenstate properties of the detuned PXP model. (a) Overlaps of all eigenstates of the PXP model with the polarized state $|\psi_0\rangle = |0\rangle$. (b) Bipartite entanglement entropy of the eigenstates in (a). The squares mark the previously known $\mathbb{Z}_2$ scarred eigenstates. (c) and (d) Same as (a) and (b) but for the PXP model with the static chemical potential $\mu_0 = 1.68\Omega$, approximately corresponding to the experimental value of detuning in Fig. 4. The stars denote the detuned scar eigenstates, which have high overlap with the $|0\rangle$ state as well as low entropy. All data are obtained by exact diagonalization of the PXP model on a ring with $N = 32$ sites in the zero-momentum and inversion-symmetric sector.

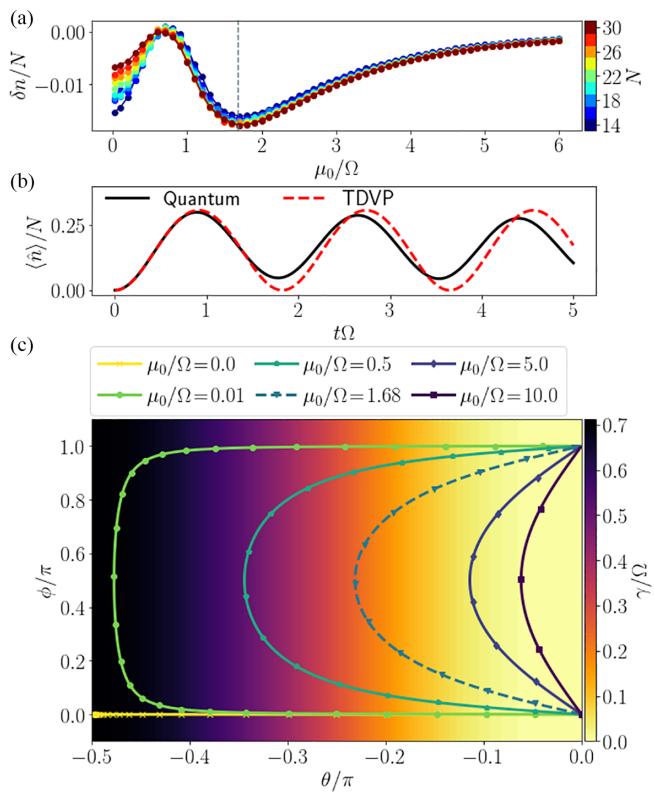

FIG. 6. Nonergodic dynamics from the polarized state in the detuned PXP model. (a) Difference between the expectation values of the diagonal and canonical ensembles for the average number of excitations. The discrepancy is maximized around $\mu_0/\Omega \approx 1.68$. (b) The average number of excitations at $\mu_0/\Omega = 1.68$ following the quench from the $|0\rangle$ state shows good agreement between the exact quantum dynamics and TDVP approximation. The exact dynamics is for system size $N = 32$ spins. (c) Trajectory in the TDVP manifold for different values of $\mu_0/\Omega$. The color scale denotes quantum leakage $\gamma$, defined in the text, which bounds the accuracy of the TDVP approximation. The markers are spaced in time by $\Delta t = 0.15/\Omega$. For the optimal value $\mu_0 = 1.68\Omega$, identified in (a) and also used in experiment, the trajectory avoids the high-leakage region and approximates well the quantum dynamics, while it is not limited to a small corner of the many-body Hilbert space.

In the single-particle case, scarred quantum dynamics originates from an unstable periodic orbit in the classical limit $\hbar \to 0$ [47]. In a many-body system, one approach to establishing a quantum-classical correspondence is to project the Schrödinger dynamics into a variational manifold, e.g., spanned by matrix product states [48], a method known as the "time-dependent variational principle" (TDVP). It was shown that the scarred dynamics of the $|\mathbb{Z}_2\rangle$ state in the PXP model is well captured by the TDVP approach, allowing one to identify a classical orbit [16]. In Figs. 6(b) and 6(c) we utilize the TDVP approach to gain a semiclassical understanding of the detuned scarred dynamics from the $|0\rangle$ state. We parametrize the TDVP manifold using translation-invariant, spin-coherent states compatible with the Rydberg blockade constraint [49]. The states are defined by the Bloch sphere angles $\theta$ and $\phi$, where $\sin(\theta)$ is proportional to the density of excitations, while $\phi$ describes the phase. In the thermodynamic limit, we can obtain classical equations of motion for $\theta$ and $\phi$ (see Ref. [50] for a detailed derivation). Figure 6(b) demonstrates that this classical dynamical system provides an excellent approximation of the quantum trajectory for sufficiently large values of $\mu_0$, including $\mu_0 = 1.68\Omega$.

To quantify the accuracy of the TDVP approach in capturing the quantum dynamics, we use "quantum leakage": the instantaneous norm of a component of the state vector that lies outside the TDVP manifold, $\gamma^2 \equiv (1/N)|| |\dot{\psi}\rangle - i\hat{H}|\psi\rangle ||^2$ [16]. For the initial state $|0\rangle$, the leakage has a simple analytic expression $\gamma^2 = \Omega^2 \sin^6\theta/(1 + \sin^2\theta)$ [50]. The leakage is higher as $\theta$ is increased, corresponding to a larger density of excitations. In this regime, i.e., for small values of $\mu_0/\Omega$, the PXP constraint has a strong effect, and the spin-coherent state ansatz does not faithfully capture the dynamics. On the other hand, for large values of $\mu_0/\Omega$, the leakage is low, but $\theta$ is confined to values near zero; thus the trajectory does not explore much of the Hilbert space. This corresponds to the trivial case where the dynamics is confined to very low densities of excitations, rendering the constraint unimportant. Finally, in the intermediate regime of $\mu_0/\Omega$ where we observe the scarring, the TDVP dynamics is able to "avoid" the high-leakage area, as seen in Fig. 6(c), while at the same time $\theta$ is not pinned to zero and the dynamics is not confined to one corner of the Hilbert space.

## VI. DISCUSSION AND OUTLOOK

We performed a quantum simulation of the paradigmatic PXP model of many-body scarring using a tilted Bose-Hubbard optical lattice. We demonstrated the existence of persistent quantum revivals from the $|\mathbb{Z}_2\rangle$ initial state and their dynamical stabilization, opening up a route for the investigation of scarring beyond Rydberg atom arrays. By harnessing the effect of detuning, we observed a scarring regime associated with the polarized initial state. As the latter state is spatially homogeneous, its preparation does not require a superlattice, which makes further investigations of scarring phenomena accessible to a large class of ultracold-atom experiments.

Moreover, we have demonstrated that periodic driving can lead to a striking decoupling of the scarred subspace from the rest of the thermalizing bulk of the spectrum, as revealed by the arrested growth of entanglement entropy. The mechanism of this enhancement is a subject of ongoing investigations. On the one hand, Ref. [51] used a kicked toy model to argue that the scarring enhancement originates from a discrete time crystalline order. On the other hand, Ref. [52] studied the cosine drive employed in experiment, finding two distinct regimes with long-lived scarred revivals. In the regime corresponding to the parameter values in Fig. 2 above, the driving parameters need to be fine-tuned to match the intrinsic revival frequency of the undriven scarred system. Moreover, the stabilization was no longer possible when the system was perturbed by terms which destroy scarring in the undriven case. This suggests that driving indeed acts as an enhancement mechanism, preventing dynamics from "leaking" into the thermalizing bulk.

Our demonstration of scarring in the $|0\rangle$ state highlights the importance of energy density. While the $|\mathbb{Z}_2\rangle$ has predominant support on the eigenstates in the middle of the spectrum, i.e., it constitutes an "infinite temperature" ensemble, the support of the $|0\rangle$ state is biased towards one end of the spectrum as a result of particle-hole symmetry breaking via the detuning potential. This suggests that, depending on the effective temperature, one can realize scarring from a much larger class of initial states with a suitable choice of detuning and periodic driving protocols. We illustrate this in Appendix D by simulating the quench of the chemical potential in the PXP model (see also Ref. [50]).

The versatility of optical lattice platforms allows one to directly probe the link between many-body scarring and other forms of ergodicity-breaking phenomena, such as Hilbert space fragmentation and disorder-free localization, as the latter can be conveniently studied in our setup by varying the tilt. In this context, we note that Ref. [12] has recently used the tilt potential to demonstrate Hilbert space fragmentation in the Fermi-Hubbard optical lattice. By contrast, in this paper we explored ergodicity breaking due to many-body scars occurring within a *single* fragment of the Hilbert space. While many-body scarring can be induced in the Fermi-Hubbard model by tuning to a similar resonance condition [53], the underlying mechanism is an approximate dimerization of the chain, which is conceptually different from the PXP-type scarring considered here.

In future work, it would be interesting to explore realizations of new scarring models by tuning to other resonance conditions and other types of lattices, including ladders and two-dimensional arrays. Indeed, it is known that the U(1) quantum link model (QLM) [54,55] can be exactly mapped to the PXP model [56]. As such, recent large-scale experiments realizing the U(1) QLM [57,58] can in principle also probe our results. A proposal has recently been introduced to extend these setups to (2 + 1)D [59], where a mapping between the U(1) QLM and PXP model does not hold, which would allow one to probe how the scarring regimes discovered in this paper would behave in higher spatial dimensions. Finally, the observation of long-lived quantum coherence due to scarring and its controllable enhancement via periodic modulation lays the foundation for applications such as quantum memories and quantum sensing [60,61].

## ACKNOWLEDGMENTS

We thank P. Hauke, B. Mukherjee, C. Turner, and A. Michailidis for useful discussions. The experiment is supported by NNSFC Award No. 12125409, the Anhui Initiative in Quantum Information Technologies, and the Chinese Academy of Sciences. A.H., J.-Y.D., and Z.P. acknowledge support from EPSRC Grant No. EP/R513258/1 and from Leverhulme Trust Research Leadership Award No. RL-2019-015. A.H. acknowledges funding provided by the Institute of Physics Belgrade, through a grant from the Ministry of Education, Science, and Technological Development of the Republic of Serbia. Part of the numerical simulations were performed at the Scientific Computing Laboratory, National Center of Excellence for the Study of Complex Systems, Institute of Physics Belgrade. J.C.H. acknowledges support from Provincia Autonoma di Trento, the ERC starting grant StrEnQTh (Project No. 804305), the Google Research Scholar award ProGauge, and Q@TN–Quantum Science and Technology in Trento. B.Y. acknowledges support from National Key R&D Program of China (Grant No. 2022YFA1405800) and NNSFC (Grant No. 12274199).

## APPENDIX A: MAPPING THE TILTED 1D BOSE-HUBBARD ONTO THE PXP MODEL AT $\Delta \approx U$ RESONANCE

The Hamiltonian describing our 1D Bose-Hubbard model is given in Eq. (2) of the main text, with $J$ denoting the hopping amplitude, $\hat{H}_U$ denoting the corresponding interaction term, and $\hat{H}_\Delta$ denoting the tilt potential. We denote by $L$ the number of lattice sites and assume open boundary conditions (OBCs). Unless specified otherwise, we fix the filling factor to $\nu = 1$, i.e., the number of bosons is equal to the number of sites in the chain.

In the $U, \Delta \gg J$ limit, the energy spectrum of the Hamiltonian in Eq. (2) splits into bands with approximately constant expectation value of the diagonal terms, $\langle \hat{H}_U + \hat{H}_\Delta \rangle \approx$ const, and the Hilbert space becomes fragmented. At the $U \approx \Delta \gg J$ resonance, the only process which conserves $\langle \hat{H}_U + \hat{H}_\Delta \rangle$ is $11 \leftrightarrow 20$, i.e., doublons can only be created by moving a particle to the left and destroyed by moving a particle to the right. In the connected component of the Fock state $|111 \cdots 111\rangle$, the system in the resonant regime is described by an effective Hamiltonian

$$\hat{H}_{\text{eff}} = -J \sum_{i=1}^{L-1} (\hat{b}_i^\dagger \hat{b}_{i+1} \hat{n}_i (2-\hat{n}_i) \hat{n}_{i+1}(2-\hat{n}_{i+1}) + \text{H.c.}), \quad \text{(A1)}$$

which results from the first-order Schrieffer-Wolff transformation applied to Eq. (2) [62]. In the Supplemental Material [42] we discuss the effect of higher-order terms in the Schrieffer-Wolff transformation.

In the remainder of this Appendix, we show that the Hamiltonian (A1) is equivalent to the PXP Hamiltonian [27,28] (see also Ref. [36] for the original derivation of the mapping and a recent review [37]). The connected component of the Hilbert space contains only certain types of two-site configurations (20, 11, 12, 02, 01), while all other two-site configurations are forbidden (22, 21, 10, 00). If we consider the configuration 20 to be an excitation, all allowed configurations can be mapped to those of the PXP model as follows:

$$\begin{aligned}
\cdots 20 \cdots &\leftrightarrow \circ \bullet \circ, \\
\cdots 11 \cdots &\leftrightarrow \circ \circ \circ, \\
\cdots 12 \cdots &\leftrightarrow \circ \circ \bullet, \quad \text{(A2)} \\
\cdots 02 \cdots &\leftrightarrow \bullet \circ \bullet, \\
\cdots 01 \cdots &\leftrightarrow \bullet \circ \circ.
\end{aligned}$$

Note that excitations live on the bonds between sites and this mapping also includes links to the two surrounding sites. For example, the configuration $\cdots 2020 \cdots$ maps to $\circ \bullet \circ \bullet \circ$ and not to $\circ \bullet \circ \circ \bullet \circ$. On the other hand, the configuration 2020 with OBCs on both sides maps to $\bullet \circ \bullet$, as there are no bonds across the boundaries.

The effective Hamiltonian (A1) can be rewritten as

$$\hat{H}_{\text{eff}} = -J \sum_{i=1}^{L-1} \left( \underbrace{\hat{b}_i^\dagger \hat{b}_{i+1} \delta_{\hat{n}_i,1} \delta_{\hat{n}_{i+1},1}}_{\sqrt{2}\hat{P}_{j-1}\hat{\sigma}_j^+ \hat{P}_{j+1}} + \underbrace{\hat{b}_{i+1}^\dagger \hat{b}_i \delta_{\hat{n}_i,2} \delta_{\hat{n}_{i+1},0}}_{\sqrt{2}\hat{P}_{j-1}\hat{\sigma}_j^- \hat{P}_{j+1}} \right). \quad (A3)$$

In this equation, the index $i$ labels the sites, while $j$ labels the bonds between sites. The Kronecker delta functions have been expressed in terms of projectors, $\hat{P}_j = |\circ_j\rangle\langle\circ_j|$, and the bosonic hopping terms correspond to the spin raising and lowering operators, $\hat{\sigma}_j^\pm$, on the bond $j$. We can use delta functions because there are no configurations with more than two particles per site in this connected component and the only possible values of $\hat{n}_i(2-\hat{n}_i)$ are 0 and 1. Moving a particle to the neighboring site on the left corresponds to creating an excitation, and moving to the right corresponds to annihilating, while delta functions act as constraints.

Finally, the effective Hamiltonian is equivalent to the PXP Hamiltonian

$$\hat{H}_{\text{PXP}} = \Omega \sum_{j=1}^{N} (\hat{P}_{j-1} \hat{\sigma}_j^+ \hat{P}_{j+1} + \hat{P}_{j-1} \hat{\sigma}_j^- \hat{P}_{j+1})$$

$$= \Omega \sum_{j=1}^{N} \hat{P}_{j-1} \hat{X}_j \hat{P}_{j+1}, \quad (A4)$$

when we set $\Omega = -\sqrt{2}J$ and $N = L - 1$, with $\hat{X}_j \equiv |\circ_j\rangle\langle\bullet_j| + |\bullet_j\rangle\langle\circ_j|$ denoting the usual Pauli $x$ matrix. In the case of OBCs, the two boundary terms become $\hat{X}_1 \hat{P}_2$ and $\hat{P}_{N-1} \hat{X}_N$. Note that the effective bosonic model in Eq. (A3) for a system size $L$ is equivalent to the PXP model for size $N = L - 1$ since the number of bonds is the number of sites minus 1.

In the PXP model, the initial states which lead to pronounced quantum revivals are the two states with the maximal number of excitations: the Néel states, $|\bullet\circ\bullet\circ\cdots\bullet\circ\rangle$ and $|\circ\bullet\circ\bullet\cdots\circ\bullet\rangle$ [23,31]. The equivalent states in the tilted Bose-Hubbard model are $|2020\cdots 201\rangle$ and $|12020\cdots 20\rangle$, for odd system sizes, and $|2020\cdots 20\rangle$ and $|120\cdots 201\rangle$ for even sizes. In our experimental setup, it is not possible to exactly prepare the $|2020\cdots 201\rangle$ state due to the inability to independently control single sites. Instead, our experiment realizes the $|2020\cdots 20\rangle$ state, which corresponds to the Néel state $|\bullet\circ\bullet\circ\cdots\bullet\circ\bullet\rangle$ in the PXP model with an odd number of sites and OBCs.

Figure 7 numerically demonstrates the mapping between the tilted Bose-Hubbard model in Eq. (2) and the PXP model in Eq. (1) in a lattice size $L = 9$. The figure shows the overlap of eigenstates with the Néel state as a function of energy, for the choice of parameters $U = \Delta = 12$ and $J = 1$. The energy spectrum is split into bands with approximately constant expectation value of the sum of interaction and tilt terms $\langle \hat{H}_U + \hat{H}_\Delta \rangle$, as indicated by different colors. The inset shows the top part of the highest-overlap band, around the energy $E = \langle 202020201|\hat{H}|202020201\rangle = 432$. This band is described by the effective Hamiltonian (A1), which preserves the expectation value $\langle \hat{H}_U + \hat{H}_\Delta \rangle$ and is equivalent to the PXP Hamiltonian. A band of scarred eigenstates is magnified in the inset, and indeed resembles similar plots for the PXP model [31]. As the two Néel states have the maximal number of doublons at filling factor $\nu = 1$, this type of dynamics also leads to oscillations in doublon number, which was experimentally measured in Fig. 2.

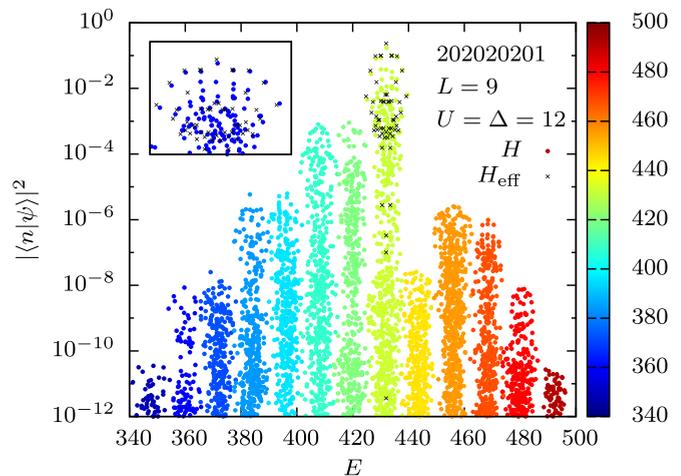

FIG. 7. Numerical demonstration of the mapping between the PXP and tilted Bose-Hubbard models. The overlap of the state $|202020201\rangle$ with the eigenstates of the tilted Bose-Hubbard model in Eq. (2) for $J = 1$ and $U = \Delta = 12$ (in units of $\hbar = 1$) is shown. The color indicates the expectation value of the diagonal part of the Hamiltonian, $\langle \hat{H}_U + \hat{H}_\Delta \rangle$, for each eigenstate. The black crosses correspond to the effective model in Eq. (A3), shifted by the energy $E = \langle 202020201|\hat{H}|202020201\rangle = 432$. The inset shows the top part of the band with the highest overlap, where a band of scarred eigenstates analogous to that in the PXP model can be seen.

## APPENDIX B: STATE PREPARATION AND DETECTION

Our experiment starts out with a two-dimensional Bose-Einstein condensate of $^{87}$Rb atoms prepared in the hyperfine state $|\downarrow\rangle = 5S_{1/2}|F = 1, m_F = -1\rangle$. By applying a microwave pulse, atoms can be adiabatically transferred to the state $|\uparrow\rangle = 5S_{1/2}|F = 2, m_F = -2\rangle$, which is resonant with the imaging laser and thus can be detected. The atoms are initially confined to a single layer of a pancake-shaped trap with 3 μm period. In both $x$ and $y$ directions, we have an optical superlattice that can be controlled separately. Each superlattice potential is generated by superimposing two standing waves with laser frequency $\lambda_s = 767$ nm and $\lambda_l = 1534$ nm, which can be described by

$$V(x) = V_s^x \cos^2(kx) - V_l^x \cos^2(kx/2 + \theta_x),$$
$$V(y) = V_s^y \cos^2(ky) - V_l^y \cos^2(ky/2 + \theta_y), \quad (B1)$$

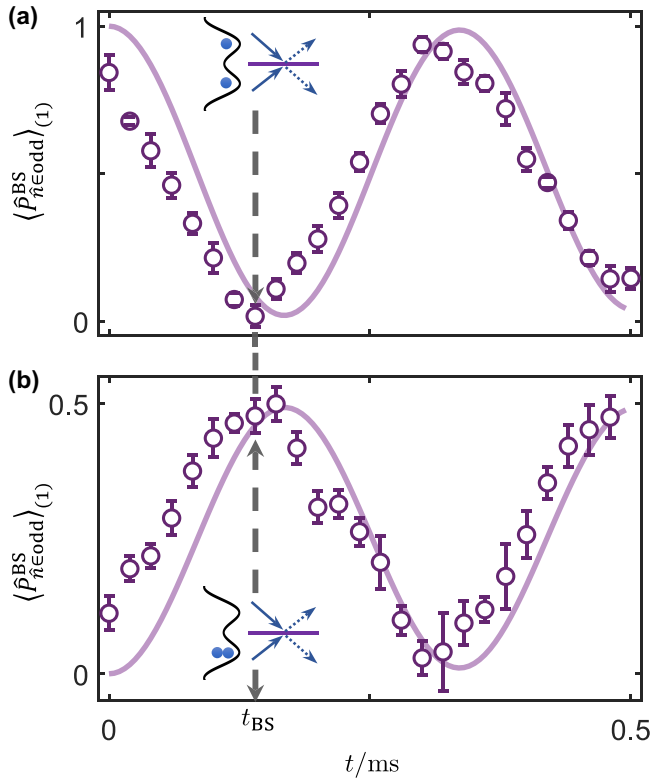

FIG. 8. Quantum interference. (a) Interfering $|1, 1\rangle$ product states in the double wells. (b) Interfering $|2, 0\rangle$ product states in the double wells. Solid curves are TEBD simulations. Experimental data are shifted forward at earliest times due to the 50 μs ramping time of the lattice potential.

where $V_{s(l)}^{x(y)}$ is the depth of the short (long) lattice in the $x$ ($y$) direction, $k = 2\pi/\lambda_s$ is the short-lattice wave number, and $\theta_{x(y)}$ is the relative phase between the short and long lattices in the $x$ ($y$) direction.

We first perform a cooling technique by loading the atoms into a staggered superlattice in the $y$ direction at $\theta_y = \pi/4$, meanwhile ramping up only the short lattice in the $x$ direction. We tune the $y$-superlattice potential to create a Mott insulator with $\bar{n} = 2$ filling in odd sites, while even sites form a $\bar{n} = 1.5$ superfluid, serving as a reservoir for carrying away the thermal entropy [34].

Atoms at even sites are removed by performing site-selective addressing. This is done by first setting $\theta_y = 0$ to form double wells and then tuning the polarization of the short-lattice laser along the $y$ direction to create an energy splitting between even and odd sites for the $|\downarrow\rangle$-to-$|\uparrow\rangle$ transition. We transfer the atoms at even sites to $|\uparrow\rangle$ and remove them with the imaging laser [35]. In this way we have prepared the initial $|\mathbb{Z}_2\rangle$ state $|2020\cdots\rangle$. The same site-selective addressing procedure is also utilized to read out atomic density on even and odd sites separately in experiment. Inside each isolated double-well unit, we can perform state engineering that transfers the state $|2, 0\rangle$ to $|1, 1\rangle$ [34]. This results in the unit-filling state $|1111\cdots\rangle$ which corresponds to the polarized state $|0\rangle$ in the PXP model.

## APPENDIX C: QUANTUM INTERFERENCE IN THE DOUBLE WELLS

The beam splitter (BS) interference is realized in the balanced double wells formed by the superlattices in the $x$ direction, expressed in Eq. (B1) by setting $\theta_x = 0$. In the noninteracting limit, indistinguishable bosonic particles coming into the interference at $t = 0$ interfere according to the bosonic bunching. Therefore having equal numbers of atoms coming into the two ports at $t = 0$ results in $\langle \hat{P}^{\text{BS}}_{\hat{n}\in \text{odd}}\rangle = 0$ at $t_{\text{BS}}$, while having different numbers of atoms interfering results in $\langle \hat{P}^{\text{BS}}_{\hat{n}\in \text{odd}}\rangle = 0.5$. Each copy of atoms coming into the interference is prepared individually, and hence there is no global phase between them, resulting in the equivalence between the two output ports [44].

To implement the quantum-interference protocol, we quench the $x$-lattice potentials to $V_s^x = 6E_r$ and $V_l^x = 5E_r$, resulting in the intra-double-well tunneling at $J \approx 740$ Hz and inter-double-well tunneling $J' \approx 35$ Hz. Simultaneously, we lower the lattice depth in the $x$ direction to $25E_r$ and trapping frequency in the $z$ direction to 1.4 kHz, achieving an interaction of $U \approx 360$ Hz. Two examples are shown here in Fig. 8, where we interfere product states $|1, 1\rangle$ [Fig. 8(a)] or $|2, 0\rangle$ [Fig. 8(b)] in the double wells and read out the average odd particle density. At $t_{\text{BS}} = 0.14$ ms we identify the beam splitter operation, where $|1, 1\rangle$ gives $\langle \hat{P}^{\text{BS}}_{\hat{n}\in\text{odd}}\rangle_{(1)} = 0.01(3)$, while $|2, 0\rangle$ gives $\langle \hat{P}^{\text{BS}}_{\hat{n}\in\text{odd}}\rangle_{(1)} = 0.48(3)$. We simulate the interference dynamics with a 20-site chain consisting of ten double-well units. We find good agreement at later times, while the earlier times are affected by the finite time in the lowering and raising of lattice potentials, which takes 50 μs. The finite interaction strength and inter-double-well tunneling together result in about 1% error in the beam splitter operation in the simulation, but this is beyond the precision of our absorption imaging.

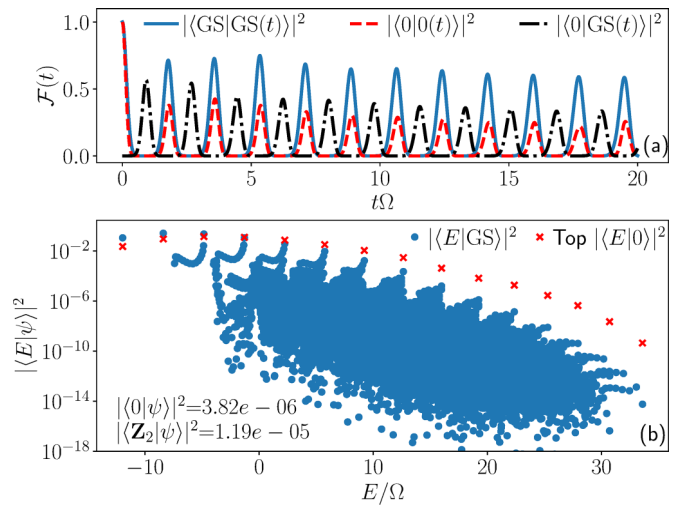

FIG. 9. Emergence of many-body scarring by quenching the chemical potential in the PXP model from $\mu_i = -0.76\Omega$ to $\mu_f = 1.6\Omega$. (a) The dynamics of quantum fidelity (blue solid curve) is similar to that of the polarized state for the same value of $\mu_f$ (red dashed curve). The overlap between the time-evolved state and $|0\rangle$ (black dash-dotted curve) shows that a significant state transfer occurs between them. (b) The overlap of the prequench ground state with the eigenstates of $\hat{H}(\mu_f)$ displays characteristic scar tower structures. Red crosses denote the highest overlaps with the $|0\rangle$ state in each scarred tower. The overlap of the prequench ground state with $|0\rangle$ and $|\mathbb{Z}_2\rangle$ states is given in the inset. All data are for $N = 32$ spins in the zero-momentum, inversion-symmetric sector of the Hilbert space.

## APPENDIX D: OTHER SCARRED STATES

In addition to the $|\mathbb{Z}_2\rangle$ and $|0\rangle$ states, we find other reviving states in the PXP model with static detuning, $\hat{H}(\mu)$, introduced in Sec. V. These initial states are the ground states of $\hat{H}(\mu_i)$, and they exhibit revivals when the detuning is quenched to a different value, $\hat{H}(\mu_i) \to \hat{H}(\mu_f)$. This setup generalizes the quench protocols studied in the main text. For example, setting $\mu_i \to -\infty$, the prequench ground state is simply the $|\mathbb{Z}_2\rangle$ state, and then quenching to $\mu_f = 0$ (pure PXP model) gives rise to scarred many-body revivals. Conversely, if we set $\mu_i \to \infty$, the ground state is $|0\rangle$, and quenching to $\mu_f = 1.68\Omega$ also leads to scarring, as this value corresponds to the Bose-Hubbard detuning value in Fig. 4.

We numerically identify similar scarring phenomenology in a larger set of initial conditions by varying the parameters $\mu_i$ and $\mu_f$. In Fig. 9 we present an illustrative example for $\mu_i = -0.76\Omega$ and $\mu_f = 1.6\Omega$. Unlike the $|\mathbb{Z}_2\rangle$ and $|0\rangle$ states, the ground state of $\hat{H}(\mu_i)$, for general values of $|\mu_i| < 2$, is not a product state. Nevertheless, such ground states have low entanglement entropy and can be prepared experimentally, while at the same time they are nearly orthogonal to the $|\mathbb{Z}_2\rangle$ and $|0\rangle$ states (the overlap with the latter is on the order $10^{-5}$). We emphasize that this does not require fine-tuning: We find large regions of $\mu_i$ and $\mu_f$ leading to scarring.

The dynamics in Fig. 9 is similar to that of the polarized state evolved with $\hat{H}(\mu = 1.68\Omega)$. During the evolution, the state periodically transfers to the polarized state and then returns to itself. The frequency of revivals is approximately the same as that for the polarized state evolved with the same static detuning $\mu_f$, but the revivals are more prominent. The overlap of the $\hat{H}(\mu_i)$ ground state with all the eigenstates of $\hat{H}(\mu_f)$ is shown in Fig. 9(b). These overlaps exhibit a similar pattern to the overlap of eigenstates with the polarized states (red crosses). Furthermore, the atypical eigenstates appear to be the same in the two cases, up to a difference in phase. This is similar to what we find for the $|\mathbb{Z}_2\rangle$ and $|\bar{\mathbb{Z}}_2\rangle$ states at $\mu_f = 0$: Both states have the same magnitude of overlap with each eigenstate, while the phases are different.

[1] J. M. Deutsch, Quantum statistical mechanics in a closed system, Phys. Rev. A **43**, 2046 (1991).

[2] M. Srednicki, Chaos and quantum thermalization, Phys. Rev. E **50**, 888 (1994).

[3] M. Rigol, V. Dunjko, and M. Olshanii, Thermalization and its mechanism for generic isolated quantum systems, Nature (London) **452**, 854 (2008).

[4] T. Kinoshita, T. Wenger, and D. S. Weiss, A quantum Newton's cradle, Nature (London) **440**, 900 (2006).

[5] R. Nandkishore and D. A. Huse, Many-body localization and thermalization in quantum statistical mechanics, Annu. Rev. Condens. Matter Phys. **6**, 15 (2015).

[6] D. A. Abanin, E. Altman, I. Bloch, and M. Serbyn, Colloquium: Many-body localization, thermalization, and entanglement, Rev. Mod. Phys. **91**, 021001 (2019).

[7] M. Serbyn, D. A. Abanin, and Z. Papić, Quantum many-body scars and weak breaking of ergodicity, Nat. Phys. **17**, 675 (2021).

[8] S. Moudgalya, B. A. Bernevig, and N. Regnault, Quantum many-body scars and Hilbert space fragmentation: A review of exact results, Rep. Prog. Phys. **85**, 086501 (2022).

[9] H. Bernien, S. Schwartz, A. Keesling, H. Levine, A. Omran, H. Pichler, S. Choi, A. S. Zibrov, M. Endres, M. Greiner, V. Vuletić, and M. D. Lukin, Probing many-body dynamics on a 51-atom quantum simulator, Nature (London) **551**, 579 (2017).

[10] D. Bluvstein, A. Omran, H. Levine, A. Keesling, G. Semeghini, S. Ebadi, T. T. Wang, A. A. Michailidis, N. Maskara, W. W. Ho, S. Choi, M. Serbyn, M. Greiner, V. Vuletić, and M. D. Lukin, Controlling quantum many-body dynamics in driven Rydberg atom arrays, Science **371**, 1355 (2021).

[11] W. Kao, K.-Y. Li, K.-Y. Lin, S. Gopalakrishnan, and B. L. Lev, Topological pumping of a 1D dipolar gas into strongly correlated prethermal states, Science **371**, 296 (2021).

[12] S. Scherg, T. Kohlert, P. Sala, F. Pollmann, B. Hebbe Madhusudhana, I. Bloch, and M. Aidelsburger, Observing non-ergodicity due to kinetic constraints in tilted Fermi-Hubbard chains, Nat. Commun. **12**, 4490 (2021).

[13] P. N. Jepsen, Y. K. Lee, H. Lin, I. Dimitrova, Y. Margalit, W. W. Ho, and W. Ketterle, Catching Bethe phantoms and quantum many-body scars: Long-lived spin-helix states in Heisenberg magnets, Nat. Phys. **18**, 899 (2022).

[14] S. Moudgalya, N. Regnault, and B. A. Bernevig, Entanglement of exact excited states of Affleck-Kennedy-Lieb-Tasaki models: Exact results, many-body scars, and violation of the strong eigenstate thermalization hypothesis, Phys. Rev. B **98**, 235156 (2018).

[15] N. Shiraishi and T. Mori, Systematic Construction of Counterexamples to the Eigenstate Thermalization Hypothesis, Phys. Rev. Lett. **119**, 030601 (2017).

[16] W. W. Ho, S. Choi, H. Pichler, and M. D. Lukin, Periodic Orbits, Entanglement, and Quantum Many-Body Scars in Constrained Models: Matrix Product State Approach, Phys. Rev. Lett. **122**, 040603 (2019).

[17] D. K. Mark, C.-J. Lin, and O. I. Motrunich, Unified structure for exact towers of scar states in the Affleck-Kennedy-Lieb-Tasaki and other models, Phys. Rev. B **101**, 195131 (2020).

[18] S. Sugiura, T. Kuwahara, and K. Saito, Many-body scar state intrinsic to periodically driven system, Phys. Rev. Res. **3**, L012010 (2021).

[19] K. Mizuta, K. Takasan, and N. Kawakami, Exact Floquet quantum many-body scars under Rydberg blockade, Phys. Rev. Res. **2**, 033284 (2020).

[20] B. Mukherjee, S. Nandy, A. Sen, D. Sen, and K. Sengupta, Collapse and revival of quantum many-body scars via Floquet engineering, Phys. Rev. B **101**, 245107 (2020).

[21] I. Mondragon-Shem, M. G. Vavilov, and I. Martin, Fate of quantum many-body scars in the presence of disorder, PRX Quantum **2**, 030349 (2021).

[22] N. Shibata, N. Yoshioka, and H. Katsura, Onsager's Scars in Disordered Spin Chains, Phys. Rev. Lett. **124**, 180604 (2020).

[23] C. J. Turner, A. A. Michailidis, D. A. Abanin, M. Serbyn, and Z. Papić, Weak ergodicity breaking from quantum many-body scars, Nat. Phys. **14**, 745 (2018).

[24] C.-J. Lin and O. I. Motrunich, Exact Quantum Many-Body Scar States in the Rydberg-Blockaded Atom Chain, Phys. Rev. Lett. **122**, 173401 (2019).

[25] T. Iadecola, M. Schecter, and S. Xu, Quantum many-body scars from magnon condensation, Phys. Rev. B **100**, 184312 (2019).

[26] V. Khemani, C. R. Laumann, and A. Chandran, Signatures of integrability in the dynamics of Rydberg-blockaded chains, Phys. Rev. B **99**, 161101(R) (2019).

[27] P. Fendley, K. Sengupta, and S. Sachdev, Competing density-wave orders in a one-dimensional hard-boson model, Phys. Rev. B **69**, 075106 (2004).

[28] I. Lesanovsky and H. Katsura, Interacting Fibonacci anyons in a Rydberg gas, Phys. Rev. A **86**, 041601(R) (2012).

[29] B. Sun and F. Robicheaux, Numerical study of two-body correlation in a 1D lattice with perfect blockade, New J. Phys. **10**, 045032 (2008).

[30] B. Olmos, M. Müller, and I. Lesanovsky, Thermalization of a strongly interacting 1D Rydberg lattice gas, New J. Phys. **12**, 013024 (2010).

[31] C. J. Turner, A. A. Michailidis, D. A. Abanin, M. Serbyn, and Z. Papić, Quantum scarred eigenstates in a Rydberg atom chain: Entanglement, breakdown of thermalization, and stability to perturbations, Phys. Rev. B **98**, 155134 (2018).

[32] E. J. Heller, Bound-State Eigenfunctions of Classically Chaotic Hamiltonian Systems: Scars of Periodic Orbits, Phys. Rev. Lett. **53**, 1515 (1984).

[33] C. J. Turner, J.-Y. Desaules, K. Bull, and Z. Papić, Correspondence Principle for Many-Body Scars in Ultracold Rydberg Atoms, Phys. Rev. X **11**, 021021 (2021).

[34] B. Yang, H. Sun, C.-J. Huang, H.-Y. Wang, Y. Deng, H.-N. Dai, Z.-S. Yuan, and J.-W. Pan, Cooling and entangling ultracold atoms in optical lattices, Science **369**, 550 (2020).

[35] B. Yang, H.-N. Dai, H. Sun, A. Reingruber, Z.-S. Yuan, and J.-W. Pan, Spin-dependent optical superlattice, Phys. Rev. A **96**, 011602(R) (2017).

[36] S. Sachdev, K. Sengupta, and S. M. Girvin, Mott insulators in strong electric fields, Phys. Rev. B **66**, 075128 (2002).

[37] K. Sengupta, Phases and dynamics of ultracold bosons in a tilted optical lattice, in *Entanglement in Spin Chains: From Theory to Quantum Technology Applications*, edited by A. Bayat, S. Bose, and H. Johannesson (Springer, New York, 2022), pp. 425–458.

[38] F. Meinert, M. J. Mark, E. Kirilov, K. Lauber, P. Weinmann, A. J. Daley, and H.-C. Nägerl, Quantum Quench in an Atomic One-Dimensional Ising Chain, Phys. Rev. Lett. **111**, 053003 (2013).

[39] F. Meinert, M. J. Mark, E. Kirilov, K. Lauber, P. Weinmann, M. Gröbner, A. J. Daley, and H.-C. Nägerl, Observation of many-body dynamics in long-range tunneling after a quantum quench, Science **344**, 1259 (2014).

[40] J. Simon, W. S. Bakr, R. Ma, M. E. Tai, P. M. Preiss, and M. Greiner, Quantum simulation of antiferromagnetic spin chains in an optical lattice, Nature (London) **472**, 307 (2011).

[41] B. Yang, H. Sun, R. Ott, H.-Y. Wang, T. V. Zache, J. C. Halimeh, Z.-S. Yuan, P. Hauke, and J.-W. Pan, Observation of gauge invariance in a 71-site Bose–Hubbard quantum simulator, Nature (London) **587**, 392 (2020).

[42] See Supplemental Material at http://link.aps.org/supplemental/10.1103/PhysRevResearch.5.023010 for additional results and background calculations supporting the results in this paper, which contains Refs. [1,2,9,10,23,26,31,51,52,56,62–68].

[43] A. M. Kaufman, M. C. Tichy, F. Mintert, A. M. Rey, and C. A. Regal, *Advances in Atomic, Molecular and Optical Physics*, 1st ed. (Elsevier, New York, 2018), Vol. 6w7, pp. 377–427.

[44] R. Islam, R. Ma, P. M. Preiss, M. Eric Tai, A. Lukin, M. Rispoli, and M. Greiner, Measuring entanglement entropy in a quantum many-body system, Nature (London) **528**, 77 (2015).

[45] S. Choi, C. J. Turner, H. Pichler, W. W. Ho, A. A. Michailidis, Z. Papić, M. Serbyn, M. D. Lukin, and D. A. Abanin, Emergent SU(2) Dynamics and Perfect Quantum Many-Body Scars, Phys. Rev. Lett. **122**, 220603 (2019).

[46] L. D'Alessio, Y. Kafri, A. Polkovnikov, and M. Rigol, From quantum chaos and eigenstate thermalization to statistical mechanics and thermodynamics, Adv. Phys. **65**, 239 (2016).

[47] E. J. Heller, Wavepacket dynamics and quantum chaology, in *Chaos and Quantum Physics*, (North-Holland, Amsterdam, 1991), Vol. 52, pp. 547–663.

[48] J. Haegeman, J. I. Cirac, T. J. Osborne, I. Pižorn, H. Verschelde, and F. Verstraete, Time-Dependent Variational Principle for Quantum Lattices, Phys. Rev. Lett. **107**, 070601 (2011).

[49] A. A. Michailidis, C. J. Turner, Z. Papić, D. A. Abanin, and M. Serbyn, Slow Quantum Thermalization and Many-Body Revivals from Mixed Phase Space, Phys. Rev. X **10**, 011055 (2020).

[50] A. Daniel, A. Hallam, J.-Y. Desaules, A. Hudomal, G.-X. Su, J. C. Halimeh, and Z. Papić, Bridging quantum criticality via many-body scarring, arXiv:2301.03631.

[51] N. Maskara, A. A. Michailidis, W. W. Ho, D. Bluvstein, S. Choi, M. D. Lukin, and M. Serbyn, Discrete Time-Crystalline Order Enabled by Quantum Many-Body Scars: Entanglement Steering via Periodic Driving, Phys. Rev. Lett. **127**, 090602 (2021).

[52] A. Hudomal, J.-Y. Desaules, B. Mukherjee, G.-X. Su, J. C. Halimeh, and Z. Papić, Driving quantum many-body scars in the PXP model, Phys. Rev. B **106**, 104302 (2022).

[53] J.-Y. Desaules, A. Hudomal, C. J. Turner, and Z. Papić, Proposal for Realizing Quantum Scars in the Tilted 1D Fermi-Hubbard Model, Phys. Rev. Lett. **126**, 210601 (2021).

[54] S Chandrasekharan and U.-J Wiese, Quantum link models: A discrete approach to gauge theories, Nucl. Phys. B **492**, 455 (1997).

[55] U.-J. Wiese, Ultracold quantum gases and lattice systems: Quantum simulation of lattice gauge theories, Ann. Phys. (Berlin) **525**, 777 (2013).

[56] F. M. Surace, P. P. Mazza, G. Giudici, A. Lerose, A. Gambassi, and M. Dalmonte, Lattice Gauge Theories and String Dynamics in Rydberg Atom Quantum Simulators, Phys. Rev. X **10**, 021041 (2020).

[57] Z.-C. Yang, F. Liu, A. V. Gorshkov, and T. Iadecola, Hilbert-Space Fragmentation from Strict Confinement, Phys. Rev. Lett. **124**, 207602 (2020).

[58] Z.-Y. Zhou, G.-X. Su, J. C. Halimeh, R. Ott, H. Sun, P. Hauke, B. Yang, Z.-S. Yuan, J. Berges, and J.-W. Pan, Thermalization dynamics of a gauge theory on a quantum simulator, Science **377**, 311 (2022).

[59] J. Osborne, I. P. McCulloch, B. Yang, P. Hauke, and J. C. Halimeh, Large-scale $2+1$D U(1) gauge theory with dynamical matter in a cold-atom quantum simulator, arXiv:2211.01380.

[60] S. Dooley, Robust quantum sensing in strongly interacting systems with many-body scars, PRX Quantum **2**, 020330 (2021).

[61] J.-Y. Desaules, F. Pietracaprina, Z. Papić, J. Goold, and S. Pappalardi, Extensive Multipartite Entanglement from su(2) Quantum Many-Body Scars, Phys. Rev. Lett. **129**, 020601 (2022).

[62] S. Bravyi, D. P. DiVincenzo, and D. Loss, Schrieffer–Wolff transformation for quantum many-body systems, Ann. Phys. (Amsterdam) **326**, 2793 (2011).

[63] J. C. Halimeh, R. Ott, I. P. McCulloch, B. Yang, and P. Hauke, Robustness of gauge-invariant dynamics against defects in ultracold-atom gauge theories, Phys. Rev. Res. **2**, 033361 (2020).

[64] B. Mukherjee, A. Sen, D. Sen, and K. Sengupta, Dynamics of the vacuum state in a periodically driven Rydberg chain, Phys. Rev. B **102**, 075123 (2020).

[65] Z. Yao, L. Pan, S. Liu, and H. Zhai, Quantum many-body scars and quantum criticality, Phys. Rev. B **105**, 125123 (2022).

[66] V. Oganesyan and D. A. Huse, Localization of interacting fermions at high temperature, Phys. Rev. B **75**, 155111 (2007).

[67] G. Vidal, Efficient Classical Simulation of Slightly Entangled Quantum Computations, Phys. Rev. Lett. **91**, 147902 (2003).

[68] J. Hauschild and F. Pollmann, Efficient numerical simulations with Tensor Networks: Tensor Network Python (TeNPy), SciPost Phys. Lect. Notes, 5 (2018), code available from https://github.com/tenpy/tenpy.